\documentclass[11pt]{article}
\usepackage{epsfig,dsfont}
\usepackage{subfigure}

\begin{document}

\thispagestyle{empty}
\vspace*{15mm}

\begin{center}

{\LARGE 
Worm algorithms for the 3-state Potts model \vskip1mm
with magnetic field and chemical potential }
\vskip15mm
Ydalia Delgado Mercado$^{\, a}$, 
Hans Gerd Evertz$^{\, b}$, 
Christof Gattringer$^{\, a}$  
\vskip5mm
$^a$Karl-Franzens University Graz \\
Institute for Physics\\ 
A-8010 Graz, Austria 
\vskip5mm
$^b$Technical University Graz \\ 
Institute for Theoretical and Computational Physics\\ 
A-8010 Graz, Austria
\end{center}
\vskip15mm

\begin{abstract}
We discuss worm algorithms for the 3-state Potts model with external field and
chemical potential. The complex phase problem of this system can be overcome by 
using a flux representation where the new degrees of freedom are dimer and monomer
variables. Working with this representation we discuss two different generalizations 
of the conventional Prokof'ev-Svistunov algorithm suitable for Monte Carlo simulations
of the model at arbitrary chemical potential and evaluate their performance.
\end{abstract}

\vskip20mm
\noindent
{\tt ydalia.delgado-mercado@uni-graz.at \\
evertz@tugraz.at \\
christof.gattringer@uni-graz.at}

\setcounter{page}0
\newpage
\noindent
\section{Introductory remarks}
 
Over the last two decades lattice methods have made considerable progress in  our
understanding of non-perturbative QCD, and now give rise to reliable quantitative
results. However, one of the areas where progress is painfully slow is the analysis of
QCD at finite density. The introduction of a chemical potential turns the fermion
determinant into a complex number, such that it cannot be used as a weight factor in a
Monte Carlo analysis -- an obstacle  known as complex phase problem. For true progress
with QCD thermodynamics on the lattice new algorithmic concepts will be necessary.
Often such new ideas are first developed and tested in simpler models before becoming
applicable to the full theory.

The worm algorithm \cite{worm} is a conceptually new approach which has already
been shown to be useful in some QCD related applications, in particular in strongly
coupled lattice QCD \cite{Chandrasekharan,Fromm} and in scalar \cite{scalar}
or low dimensional lattice
field theories \cite{lowdim}. In this paper we extend the range of applications
of worm algorithms to QCD related problems, and show that a worm algorithm can
be used to simulate the 3-state Potts model in three dimensions with a center
symmetry breaking term and a chemical potential, which in its standard
representation has a complex phase problem. This model has been discussed as an effective
theory for parts of the QCD phase diagram \cite{effectivemodel,flux,Alford}.
Numerical studies with various techniques for vanishing \cite{Gavai,Karsch} and
non-vanishing chemical potential \cite{Alford,Condella,Kim,Forcrand} may be
found in the literature. Recently a systematical analysis of the phase diagram as
function of temperature and chemical potential was presented and its implications
for QCD were discussed \cite{effcenter}. In this work we discuss in detail 
the worm algorithms used in \cite{effcenter} and assess their performance.

The 3-state Potts model with magnetic field $\kappa$ and chemical potential $\mu$ in 
$d$ dimensions is described by the Hamiltonian
\begin{equation}
H[P] \; = \; - \sum_{x}  \left( \tau \sum_{\nu=1}^d 
\Big[ P(x) P(x\!+\!\hat{\nu})^* + c.c. \Big] \; + \;
\eta P(x) \, + \, \overline{\eta} P(x)^* \right)\; ,
\label{action_original}
\end{equation}
where we use the variables $\eta$ and $\overline{\eta}$ which are related to the 
strength $\kappa$ of the external field and the chemical potential $\mu$ via
\begin{equation}
\eta \; = \; \kappa \, e^{\, \mu} \quad , \qquad \overline{\eta} \; = \; 
\kappa \, e^{-\mu} \; .
\end{equation}
The first sum in (\ref{action_original}) 
is over all sites $x$ of a $d$-dimensional (hyper-) cubic lattice 
with periodic boundary conditions and a total of $V$ lattice points. The second sum is 
over all 
directions $\nu = 1, 2, \, ... \, d$, where $\hat{\nu}$ denotes the unit  vector
in direction $\nu$. The spin variables $P(x)$ are complex phases with three
different values,  $P(x) \in \mathds{Z}_3 = \{1, e^{i 2 \pi / 3}, e^{-i2\pi / 3} \}$.  The coupling
parameters $\tau, \kappa$ and $\mu$ are real and non-negative. The partition function
of the model is given by $Z = \sum_{\{P\}} \, e^{-S[P]}$, where the sum runs
over all possible configurations of the phases $P(x) \in \mathds{Z}_3$.

In the context of an effective theory for QCD thermodynamics one considers the
$d=3$ model and the spin variables $P(x)$ are interpreted as static quark
sources at a spatial position $x$ (local Polyakov loops). A vanishing
expectation value $P = V^{-1} \langle \sum_x P(x) \rangle$, which we here will often
refer to as the magnetization $P$, signals confinement of quarks, while
$P \neq 0$ corresponds to deconfined quarks. The parameter
$\tau$ is a monotonically increasing function of the QCD temperature. $\kappa$
is proportional to the number of flavors and decreases with increasing quark
mass. The value $\kappa = 0$ then corresponds to pure gluodynamics. $\mu$ has
the interpretation of the quark chemical potential in units of the  inverse
temperature.

For vanishing external field, $\kappa = 0$ (i.e., $\eta = \overline{\eta} = 0$), 
the 3-state Potts model in three dimensions is known to have a first order 
transition at $\tau = 0.183522(3)$ \cite{Karsch,Janke}. For small external field
and vanishing chemical potential (i.e., $\eta = \overline{\eta} = \kappa$)
the first order transition persists, giving rise to a short first order line
which ends in a critical endpoint at $(\tau,\kappa) = (0.183127(7),0.00026(3))$
\cite{Karsch}. The case of non-vanishing chemical potential has been analyzed
with techniques based on the Swendsen-Wang cluster algorithm using improved
estimator techniques in \cite{Alford} and reweighting in
\cite{Kim}. Within the flux representation \cite{flux} also local Metropolis 
updates were used \cite{Alford,Condella}. It has  been demonstrated that
turning on the chemical potential mildens the transition and shifts the critical
endpoint towards smaller values of $\kappa$.

In this article we show that in the flux representation \cite{flux} the model is
accessible also with suitably constructed worm algorithms.  We discuss two different
types of generalized worm algorithms: Type-I: A simple generalization of the
conventional Prokof'ev-Svistunov worm \cite{worm} with steps where monomers are
inserted, combined with a random hop to a lattice site with another monomer insertion.
Type-II: Here the worms are open strings with monomers at their
endpoints (contrary to the Type-I worms which are closed but with the possibility of
intermediate hops).  

The two algorithms are compared and evaluated using bulk observables: 
the magnetization $P$, the internal
energy $U$, the susceptibility $\chi_P$ and the heat capacity $C$.

\section{Flux representation}

In this section we briefly review the flux representation \cite{flux} that we use 
for the two worm algorithms to set the notation and 
to discuss the flux representation of the observables.    
For the nearest neighbor terms of (\ref{action_original}) we use the ansatz 
\begin{equation}
e^{\, \tau [ P(x) P(x + \hat{\nu})^* + c.c.]} \;\; = \;\; C \!\!
\sum_{b_{x,\nu} = \, -1}^{+1} \,
B^{|b_{x,\nu}|} \;\, \Big( \, P(x) P(x\!+\!\hat{\nu})^* \, \Big)^{b_{x,\nu}} \; .
\label{linkfactor}
\end{equation}
In this expression the term living on the link $(x,\nu)$ is written as a 
sum over a dimer variable $b_{x,\nu} \in \{-1,0,+1\}$. A straightforward
calculation gives the constants $C$ and $B$,
\begin{equation}
C \; = \; \frac{e^{2\tau} + 2 e^{-\tau}}{3} \quad , \qquad  
B \; = \; \frac{e^{2\tau} - e^{-\tau}}{e^{2\tau} + 2 e^{-\tau}} \; . 
\end{equation}
For the magnetic field terms, which live on a single site $x$, we use a similar 
ansatz,   
\begin{equation}
e^{\, \eta \, P(x) \; + \; \overline{\eta} \, P(x)^*}  \; = \;
\sum_{s_x=-1}^{+1} M_{s_x} \, P(x)^{s_x} \; ,
\label{sitefactor}
\end{equation}
which expresses the lhs.\ as a sum over a monomer variable $s_x \in \{-1,0,+1\}$
attached to the site $x$. Again one easily works out the monomer
weights $M_s$ for $s = -1,0,+1$ and obtains
\begin{equation}
M_s   =    \frac{1}{3} \! \left[ e^{\, \eta + \overline{\eta}} \, + \, 
2 \, e^{-(\eta + \overline{\eta})/2} \, \cos\Big( (\eta - \overline{\eta} )
\frac{\sqrt{3}}{2} - s \frac{2\pi}{3} \Big) \right] .
\label{monoweights}
\end{equation}
The weights $M_s$ turn out to be non-negative for all values of 
$\eta$ and $\overline{\eta}$.

With the expressions (\ref{linkfactor}) and (\ref{sitefactor}) the partition sum 
assumes the form ($V$ again denotes the total number of lattice points)
\begin{equation}
Z \; = \; C^{\,d V} \sum_{\{b\}} \sum_{\{s\}}  
W[b,s] \, \left( \prod_x \sum_{P(x)} \, 
P(x)^{\sum_\nu[b_{x,\nu} - b_{x-\hat{\nu},\nu}] + s_x } \right) \; .
\label{step1}
\end{equation}
The first two sums in (\ref{step1}) run over all configurations of the dimer
variables $b_{x,\mu}$ and of the monomer variables $s_x$.
We have introduced the weight factor
\begin{equation}
W[b,s] \; = \; \left( \prod_{x,\nu} B^{|b_{x,\nu}|} \right) \left(
\prod_x \, M_{s_x} \right) \; .
\label{weight}
\end{equation}
The summation over the
phases $P(x)$ can now be performed in closed form. Using that the sum over the 
roots of unity vanishes, one finds ($n$ is an integer),
\begin{equation}
\sum_{P} P^{\,n} \; = \; 3 \, T(n) \quad \mbox{with} \quad T(n) \, = \, \delta_{\,n \,
\mbox{\small mod} \, 3 \, , \, 0} \; \; \; ,
\label{constraint}
\end{equation}
where we introduced the triality function $T(n)$ which equals 1 if $n$ is a
multiple of 3 and vanishes otherwise. The partition sum
in its final form reads
\begin{equation}
Z \; = \; (3 C^d)^V \sum_{\{b\}} \sum_{\{s\}}  
W[b,s] \, \prod_x  T \left( \sum_\nu [b_{x,\nu} - b_{x-\hat{\nu},\nu}] 
+ s_x \right) \; .
\label{zflux}
\end{equation} 
In (\ref{zflux}) the Potts model is represented
in terms of the dimer variables $b_{x,\nu} \in \{-1,0,+1\}$ which live on the
links $(x,\nu)$ of the lattice, and the monomer variables $s_{x} \in \{-1,0,+1\}$ 
attached to the sites $x$. Each configuration of dimers and monomers comes
with a real and non-negative weight $W[b,s]$ as given in (\ref{weight}) and thus the
complex phase problem is solved. The weight
consists of a factor $B$ for every non-vanishing dimer, and a monomer factor 
$M_{s_x}$ for each site $x$, according to the monomer variable $s_x$ at that site. 

The configurations of dimers and monomers must obey
the constraint given by the second factor in (\ref{zflux}). This factor
forces the total flux of dimers and monomers at each site $x$ 
to be a multiple of 3. 

\begin{figure}[ht]
\begin{center}
\includegraphics[width=12cm,clip]{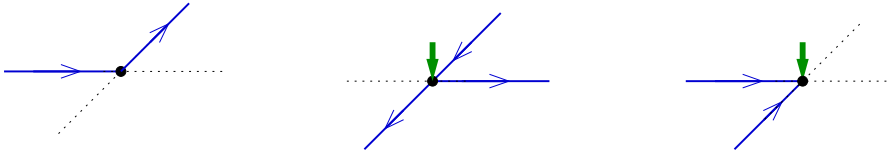}
\end{center}
\caption{Examples of admissible fluxes at a site $x$ for the case of $d=2$
dimensions. See the text for the description of the representation.}
\label{vertices}
\end{figure}

In Fig.~\ref{vertices} we show (for the 2-dimensional case) a few admissible
flux configurations at individual sites: dimers are represented by arrows in
the horizontal plane and monomers by arrows on the vertical axis. If an arrow
points towards a site (away from the site) it contributes with $+1$ ($-1$) to
the total flux at that site.  Trivial dimers, $b_{x,\nu} = 0$, and trivial
monomers, $s_x = 0$, do not contribute to the constraint and thus need not be
represented  in the illustration of admissible fluxes.

In our tests of the worm algorithms we use the bulk observables: internal
energy $U$, the magnetization $P$, i.e., the expectation value of the averaged spin variable $P = V^{-1} \sum_x
\langle P(x) \rangle$,  the heat capacity $C$ and the susceptibility $\chi_P$.
These quantities may be obtained as suitable derivatives of $\ln Z$. For the
evaluation of these derivatives it is useful to note the following relations
for derivatives of the monomer weights $M_s$:
\begin{equation}
\frac{\partial M_{+1} }{\partial \eta}  = 
\frac{\partial M_{-1}}{\partial \overline{\eta}} = M_0 \; , \;
\frac{\partial M_0 }{\partial \eta} = 
\frac{\partial M_{+1}}{\partial \overline{\eta}} = M_{-1} \; , \;
\frac{\partial M_{-1}}{\partial \eta} = 
\frac{\partial M_{0}}{\partial \overline{\eta}} = M_{+1} \; .
\end{equation}
Exploiting these relations one obtains for the expectation value of the 
spatially averaged spin the following 
representation in terms of monomers 
\begin{equation}
 P \;  \equiv \;  \frac{1}{V} \, \frac{\partial}{\partial \eta} \ln Z 
\; = \; \frac{1}{V} \left\langle 
{\cal S}_{+1} \, \frac{M_0}{M_{+1}} \, + \, 
{\cal S}_{0} \, \frac{M_{-1}}{M_0} \, + \,
{\cal S}_{-1} \, \frac{M_{+1}}{M_{-1}} \right \rangle \; ,
\label{ploopexp} 
\end{equation}
where ${\cal S}_{+1}$ denotes the number of sites $x$ where the monomer 
variables have the value $s_x = +1$, etcetera.

In a similar way one obtains for the internal energy
\begin{eqnarray}
U & \equiv & \langle H \rangle \; = \; \left[ - \tau \frac{\partial}{\partial \tau}
- \eta \frac{\partial}{\partial \eta} - 
- \overline{\eta} \frac{\partial}{\partial \overline{\eta}} \right] \ln Z \\
& = & -2 d V \tau B \, - \, \left\langle {\cal B} \,
\widetilde{B} \, + \,  
{\cal S}_{+1} \, \widetilde{M}_{+1} \, + \, 
{\cal S}_{0} \, \widetilde{M}_0 \, + \,
{\cal S}_{-1} \, \widetilde{M}_{-1} \right \rangle \; , 
\nonumber
\end{eqnarray}
with $\cal{B}$ denoting the number of links $(x,\nu)$ where the dimer variables 
are non-zero, $b_{x,\nu} \neq 0$. The new numerical constants in this
expression are
\begin{eqnarray}
\widetilde{B} & = & \frac{9 \, \tau}{e^{3\tau} + 1 - e^{-3\tau}} \quad , \quad
\widetilde{M}_{+1} \; = \; \frac{\eta M_0 + \overline{\eta}M_{-1}}{M_{+1}} \; , 
\nonumber \\ 
\widetilde{M}_{0} & = & \frac{\eta M_{-1} + \overline{\eta}M_{+1}}{M_{0}} \quad ,
\quad 
\widetilde{M}_{-1} \; = \; \frac{\eta M_{+1} + \overline{\eta}M_{0}}{M_{-1}} \; .
\end{eqnarray}
Using equivalent steps also the heat capacity $C$ and the susceptibility $\chi_P$ can be
expressed in terms of the monomer and dimer numbers and their fluctuations. For the
evaluation of spin correlators one may use locally varying weights $\eta$ and
$\overline{\eta}$ as generating functionals. The spin correlators are then 
related to correlators of monomers.

\newpage
\section{Worm algorithms}

In a worm algorithm \cite{worm}, the constraint of flux conservation
(\ref{constraint}) is temporarily relaxed during a Monte Carlo move. One unit of
flux is inserted at some lattice site and then moved through the lattice by local
steps, until the insertion is healed again, either by coming back to the
original site, or by a change of the local monomer number. The individual local
steps are performed with Metropolis update probabilities. Each complete worm
makes up one update step in the model with constraint (\ref{constraint}) which
is to be simulated. 

As outlined in the introduction, we use two different types of worm algorithms.
Type-I: The worms are closed, and in addition to the usual local worm
propagation, there are random hops with insertion of monomers. Type-II: The
worms are open strings on the lattice with monomers at the endpoints.  

\subsection{Type-I: Closed worm algorithm}
\label{cw}
\begin{figure}[b!]
\begin{center}
\vspace{-2mm}
\includegraphics[width=9cm,clip]{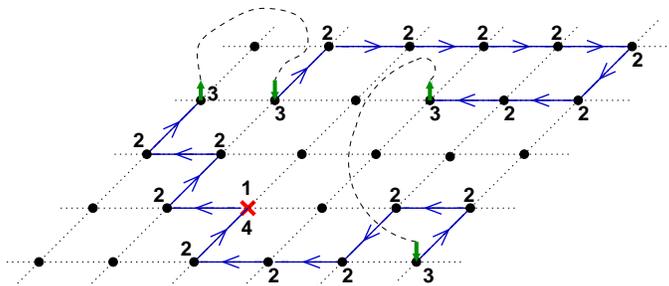}
\end{center}
\vspace{-5mm}
\caption{Example of a Type-I closed worm on an initially empty 2-D lattice.}
\label{worm2}
\end{figure}

Each worm is generated using four different moves. We illustrate them in
Fig.~\ref{worm2} for the simplest case of an initially empty lattice, i.e., all
dimer- and monomer variables are set to 0 in the beginning. The worm starts at a
random position of the lattice (1).  It may decide to insert dimers and move to the
neighboring site (2) but also monomers can be inserted (3). The insertion of a monomer is followed by a sequence of attempts to insert another monomer at a randomly chosen site, which will succeed (3) 
with a probability depending on the weights $M_s$. These steps are
continued until the worm closes (4). The acceptance of each step is governed by a
Metropolis decision. 

The subsequent pseudo-code describes the algorithm for the general case of $d$
dimensions (for the numerical tests discussed later we set $d = 3$). By $x_0$ we
denote the starting point of the worm and $x$ is the current position of its
head. $worm\_sign$ is a variable that determines  whether the worm raises or
lowers the flux. $insert\_monomer\_flag$ is a logical variable controlling the
status of the worm. By $x \oplus y$ we denote addition modulo 3 which is the
usual addition operation except in the cases $x=y=1$ and $x=y=-1$ where we
define $+1\oplus+1=-1$ and $-1\oplus-1=+1$. Finally  {\tt rand()} is a random
number generator for uniformly distributed real numbers in the interval $[0,1)$.

\vskip5mm
\noindent
\underline{{\bf Pseudocode for Type-I closed worms:}}
\vskip3mm
{\tt 
\noindent select a lattice site $x_0$ randomly

\noindent $x \, \longleftarrow \, x_0$

\noindent select $worm\_sign \in \{-1,+1\}$ randomly

\noindent $insert\_monomer\_flag$ $\; \longleftarrow \;$  false  

\vskip3mm
\noindent 
repeat until worm is complete:

if $insert\_monomer\_flag$ is true then

\hspace*{5mm} select a lattice site $x'$ randomly

\hspace*{5mm} $\widetilde{s} \, \longleftarrow \, s_{x'} \, \oplus \, (-worm\_sign)$ 

\hspace*{5mm} if rand() $ \leq \; M_{\widetilde{s}} / M_{s_{x'} }$ then

\hspace*{10mm} $s_{x'} \, \longleftarrow \, \widetilde{s}$

\hspace*{10mm}  $x \, \longleftarrow \, x'$

\hspace*{10mm} $insert\_monomer\_flag$ $\; \longleftarrow \;$  false

\hspace*{5mm} end if

else

\hspace*{5mm} select $\nu \in \{0, \pm 1, \pm 2, \, ... \, \pm d \}$ randomly

\hspace*{5mm} if $\nu = 0$ then

\hspace*{10mm} $\widetilde{s} \, \longleftarrow \, s_{x} \, \oplus \, worm\_sign$ 

\hspace*{10mm} if rand() $ \leq \; M_{\widetilde{s}} / M_{s_{x} }$ then

\hspace*{15mm} $s_{x} \, \longleftarrow \, \widetilde{s}$

\hspace*{15mm} $insert\_monomer\_flag$ $\; \longleftarrow \;$  true

\hspace*{10mm} end if

\hspace*{5mm} else

\hspace*{10mm} $\widetilde{b} \, \longleftarrow \, b_{x,\nu} \, \oplus \, (sgn(\nu)\times worm\_sign)$

\hspace*{10mm} if rand() $ \leq \; B^{\, |\widetilde{b}| \, - \, | b_{x,\nu}|}$ then

\hspace*{15mm} $b_{x,\nu} \,  \longleftarrow \, \widetilde{b}$ 

\hspace*{15mm} $x \, \longleftarrow \, x \, + \, \hat{\nu}$

\hspace*{10mm} end if

\hspace*{5mm} end if

end if

\vskip1mm

if $x = x_0$ and $insert\_monomer\_flag$ is false then

\hspace*{5mm} worm is complete

endif

\noindent
end repeat until worm is complete
}

\vskip3mm
\noindent
It is straightforward to show detailed balance using the Boltzmann weight $W[b,s]$ 
given in (\ref{weight}), and that the algorithm is ergodic.

\subsection{Type-II: Open worm algorithm} 
\label{ow} 
\begin{figure}[h!]
\begin{center}
\includegraphics[width=9cm,clip]{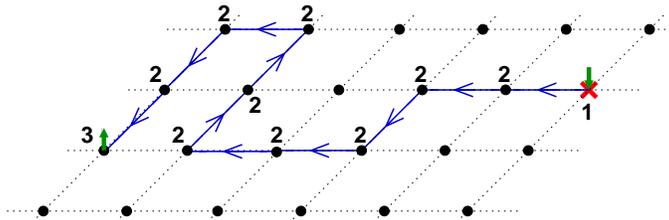}
\end{center}
\vspace{-4mm}
\caption{Example of a Type-II open worm on an initially empty 2-dimensional lattice.
See the text for its discussion.\vspace{-4mm}}
\label{worm open}
\end{figure}

We also employed a second variant of the algorithm which we refer to as Type-II open worms.
In this case, each worm is generated using three moves which we
again illustrate for an initially empty lattice (Fig.~\ref{worm open}): 
The worm tries to start at a
random position of the lattice by inserting a monomer (1). It then may decide to
insert dimers or monomers.  If a dimer is inserted the worm moves to the
neighboring site (2).  If a monomer is inserted, the worm is complete and
terminates (3). Again the acceptance of the moves is decided with a Metropolis step.
 
We stress that the Type-II open worm will not work in the case of vanishing external 
magnetic field, i.e., for $\kappa = 0$. In that case no monomers appear and the Type-II 
worm cannot start. For small $\kappa$ the Type-II open worm will have a very
small probability for starting, since the monomer weights are linear in $\kappa$ (for small
values of $\kappa$). For larger values of $\kappa$ there is no 
starting problem of the Type-II worms (see also Section 4.2).  For $\kappa \neq 0$ it is straightforward to show detailed 
balance and ergodicity for Type-II worms, and we found that they are even simpler to implement 
than the closed worms.

\section{Numerical assessment of the worms}

\begin{figure}[t!]
\centering
\includegraphics[width=0.7\textwidth,clip]{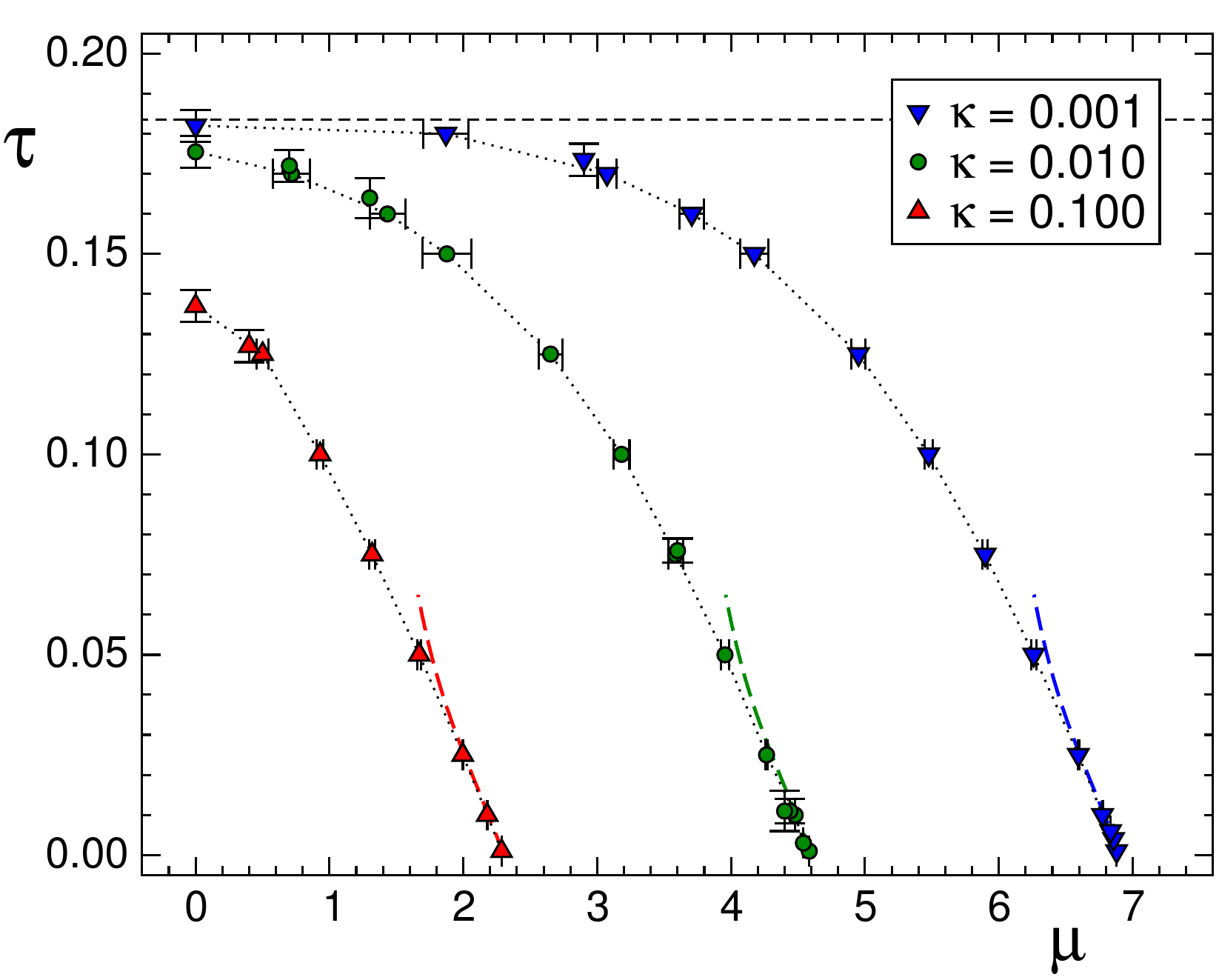}
\vspace*{-3mm}
\caption{Phase diagram \cite{effcenter} as obtained from the maxima of the
susceptibility of $P$. We show results for three values of
$\kappa$. The dashed curves at the bottom are the results of
a perturbative expansion in $\tau$ and the horizontal line marks the critical value of
$\tau$ for the $\kappa = 0$ case.} \label{phaseline}
\end{figure}

In this section we show the results of our tests and performance analysis  for the two
generalized worm algorithms. For a detailed discussion of the physics results obtained in
the simulations of the generalized Potts model we refer the reader to \cite{effcenter}.
Here we only include a figure of the phase diagram to  allow the reader to locate the
parameter values of the simulations discussed in this section in the phase diagram. In
Fig.~\ref{phaseline} we show the phase diagram in the $\tau$-$\mu$ plane obtained from
the position of the maxima of $\chi_P$  for three values of $\kappa$. The dashed curves
at the bottom are the results of a perturbative expansion in $\tau$ and the horizontal line
marks the critical value of $\tau$ for the $\kappa = 0$ case. To the left of the phase
lines the Polyakov loop is small (confinement), while it is close to 1 to the right of the phase
lines (deconfined phase). The transitions are of a crossover nature.

\subsection{Validity of the algorithms and the flux-based observables}

\begin{figure}[t!]
\centering
\includegraphics[width=90mm,clip]{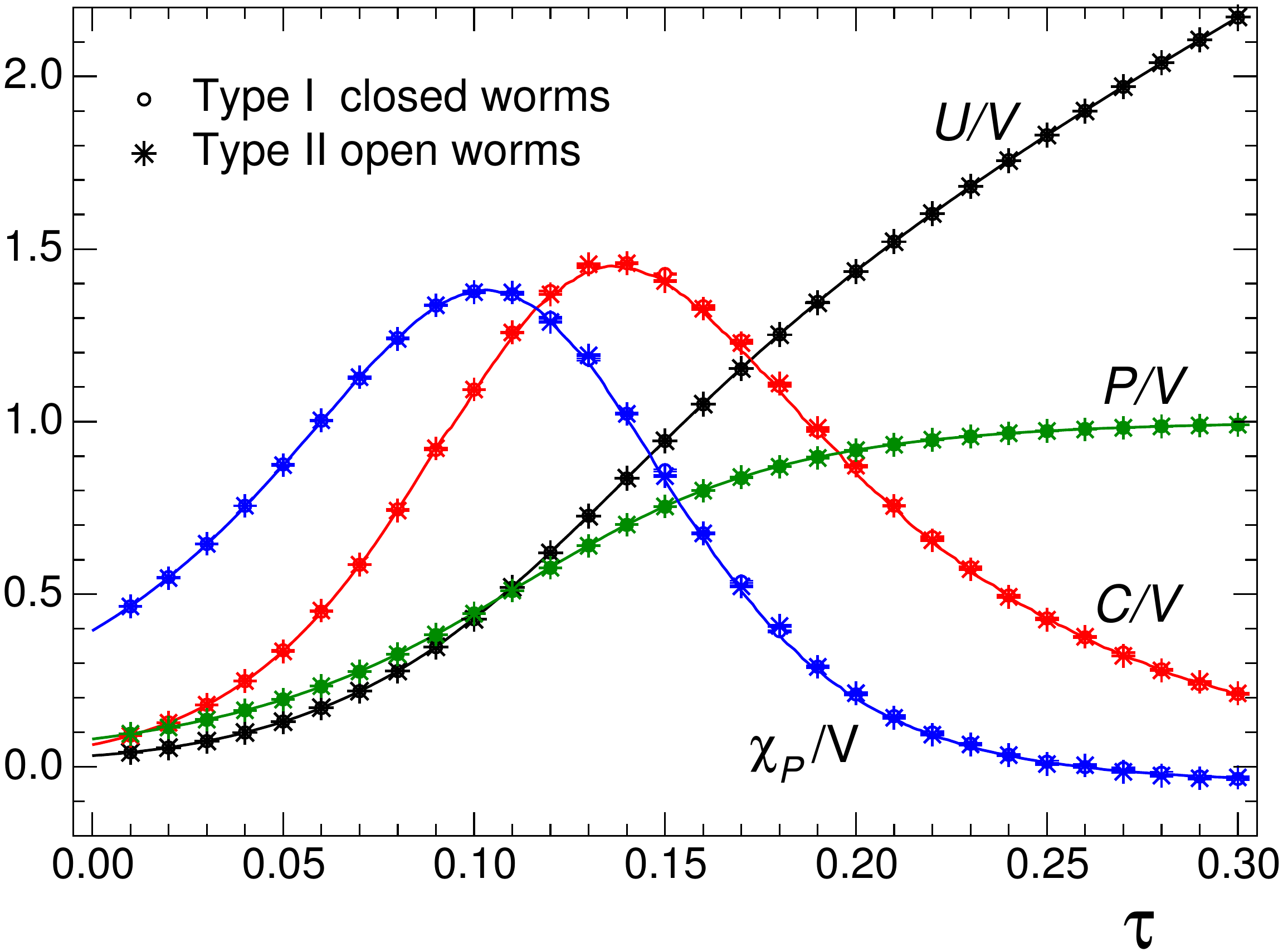}
\vspace*{-2mm}
\caption{Comparison of the exact result (full curves) on a $2 \times 2 \times 3$ lattice with the
two worm algorithms (symbols). We show the results for $U, C, P$ and $\chi_P$ as a
function of $\tau$ using $\kappa = 0.001$ and $\mu = 6.0$. Note that the data
points from the two algorithms agree almost perfectly such that the
corresponding symbols essentially fall on top of each other.} \label{smallV}
\end{figure}

As a first test we compare the results from the two worm algorithms with the results from an
exact evaluation on a small $2 \times 2 \times 3$ lattice using the original spin formulation
of the model. We consider the observables $U, C, P$ and $\chi_P$ normalized by the volume. In
Fig.~\ref{smallV} we show these observables as a function of  $\tau$ for $\kappa = 0.001$ and
$\mu = 6.0$. The curves represent the results from the  exact evaluation, while for the data
from the worm algorithms symbols are used. It is obvious, that the numbers from the exact
evaluation and from the worm algorithms agree very well. This not only establishes the validity
of the worm algorithms, but also of the mapping of the physical observables onto the flux
representation. The test on the small volume was repeated for different values of the
parameters $\tau, \kappa$ and $\mu$ and we always found excellent agreement of the worm results
with the exact evaluation of the observables.

\begin{figure}[t]
\centering
\includegraphics[width=\textwidth,clip]{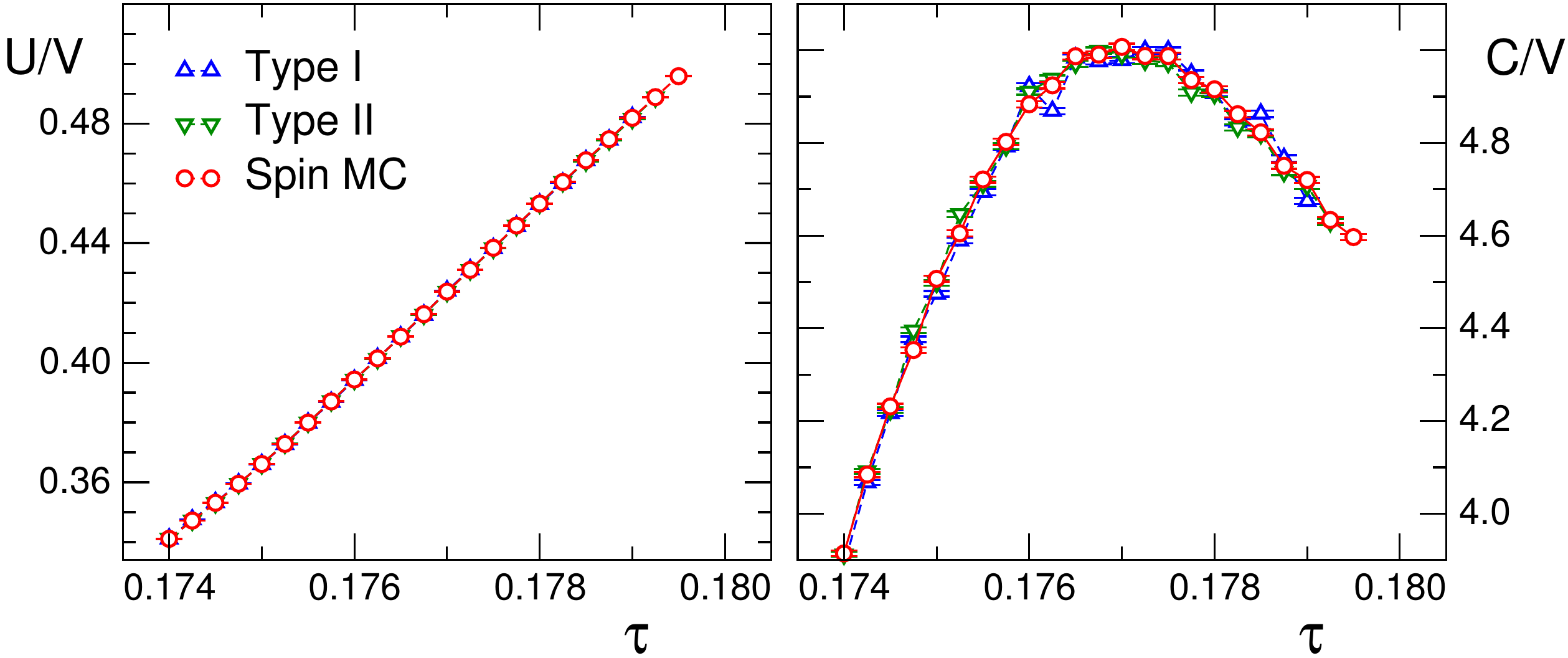}
\vspace*{-2mm}
\caption{Internal energy $U$ (left) and heat capacity $C$ (right) for $\kappa = 0.01$ and 
$\mu = 0$ on a $16^3$ lattice at values of $\tau$ near the crossover. We compare the
results from the conventional Metropolis simulation in the spin representation (red
circles), the Type-I closed worm algorithm (blue upward pointing triangles) and the
Type-II open worms (green downward pointing triangles).} \label{mu0} \end{figure}

In addition to the comparison to exact results on small volumes we performed also a
comparison with a conventional Metropolis calculation in the spin representation on 
larger volumes.  This
comparison is of course limited to vanishing chemical potential, $\mu = 0$, where the
complex phase problem is absent also in the spin representation and the conventional
Metropolis approach is possible. In Fig.~\ref{mu0} we compare the results for the
internal energy $U$ and the heat capacity $C$ as a function of $\tau$ calculated with the
three algorithms: Conventional Metropolis (red circles),  closed worm (blue upward
pointing triangles) and open worm (green downward pointing triangles)  for $\kappa =
0.01$ and $\mu = 0$ on a $16^3$ lattice for values of $\tau$ 
near the crossover. The Metropolis simulation uses $10^6$ sweeps of
local Metropolis steps for equilibration and $10^6$ measurements separated by 20 sweeps
for decorrelation. For both worm algorithms we  used $10^6$ worms for equilibration and
$10^6$ measurements. For the Type-I closed worm we  separated two measurements by $N_{deco} = 20$
worms for decorrelation, while for the Type-II open worms 5000 worms were used for
decorrelation. The results from   the three algorithms fall on top of each other. The
same tests were conducted at other values of the  parameters and also for $P$ and
$\chi_P$. Furthermore, for $\mu > 0$ where simulations with the conventional spin-based
Metropolis algorithm are not possible, we performed a comparison among the two worm
algorithms and also in this case found excellent agreement. We conclude that the two worm
algorithms are valid, and that the representation of the observables in terms of fluxes
was implemented correctly.

\subsection{Time series}
\begin{figure}[t!]
\centering
\includegraphics[width=\textwidth,clip]{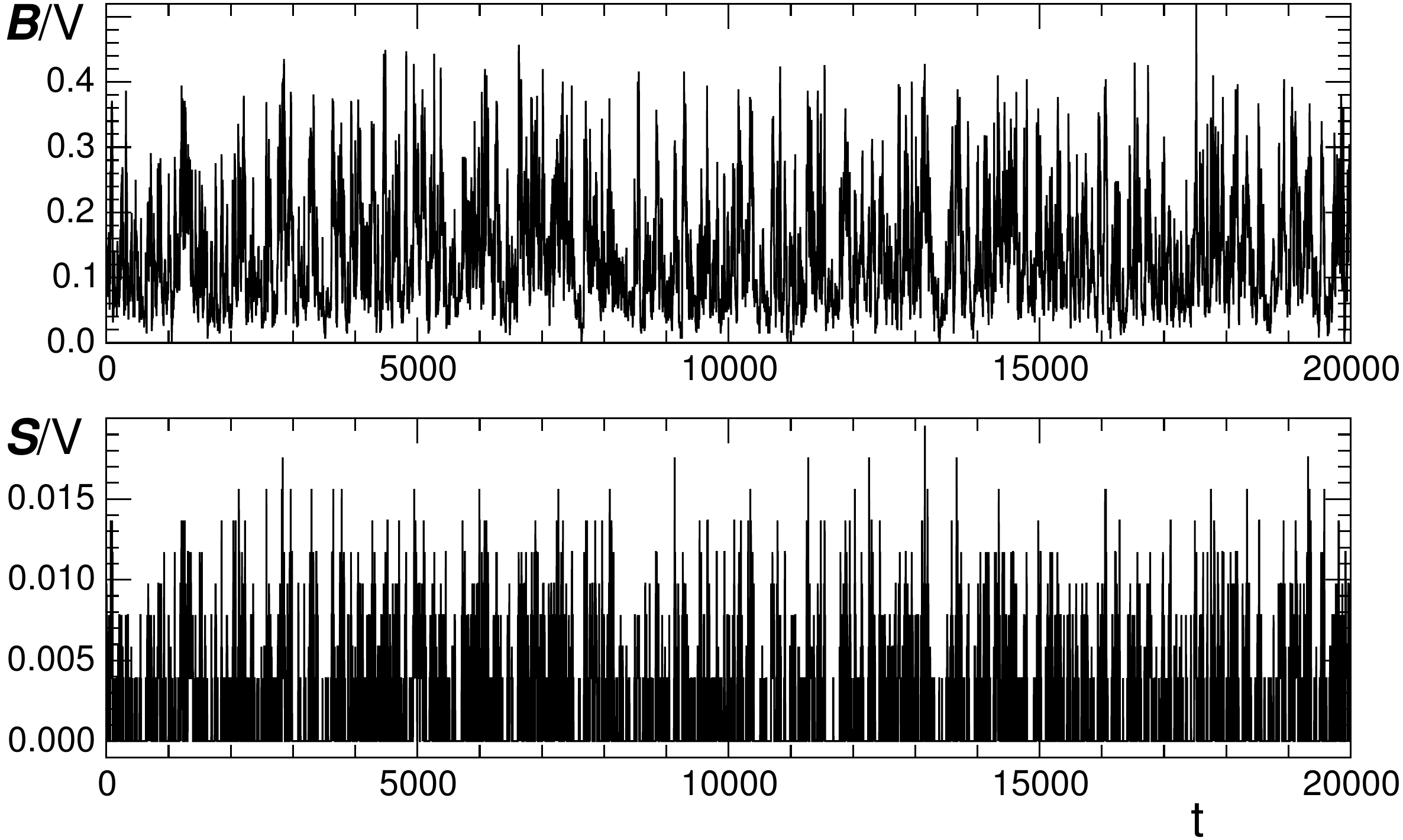}
\vspace*{-2mm}
\caption{Time series for ${\cal B}/V$ and ${\cal S}/V$ at $\tau = 0.1782$,
$\kappa = 0.005$, and $\mu = 0$ 
on a $8^3$ lattice, generated with Type-I worms. The Monte Carlo time on the horizontal
axis is measured in units of $N_{deco} = 20$ worms.} \label{large_kappa}
\end{figure}

In this subsection we now study performance aspects of the two worm algorithms. 
We begin our assessment with analyzing the time series of measurements of the
total number ${\cal B}$ of non-zero dimer variables  and the number ${\cal S}$
of non-zero monomers. As shown at the end of Section 2,  the physical
observables can be constructed from ${\cal B}$ and ${\cal S}$ and their moments. In
Fig.~\ref{large_kappa} we show time series from Type-I worms for ${\cal B}/V$ in
the  top plot and for ${\cal S}/V$ (bottom) for the parameters $\tau = 0.1782,\,
\kappa = 0.005,\, \mu = 0.0$ and a volume of $8^3$ (time series for Type-II worms
behave similarly).  Between two measurements we used $N_{deco} = 20$ worms for
decorrelation, so the Metropolis time on the horizontal axis is measured in
units of 20 worms. We do not observe structures on scales  larger than several
hundred worms and as a first impression conclude that the Type-I worms lead to
reasonably quick sampling. The sampling becomes better for larger  values of
$\kappa$ or $\mu$.

\begin{figure}[t!]
\centering
\includegraphics[width=\textwidth,clip]{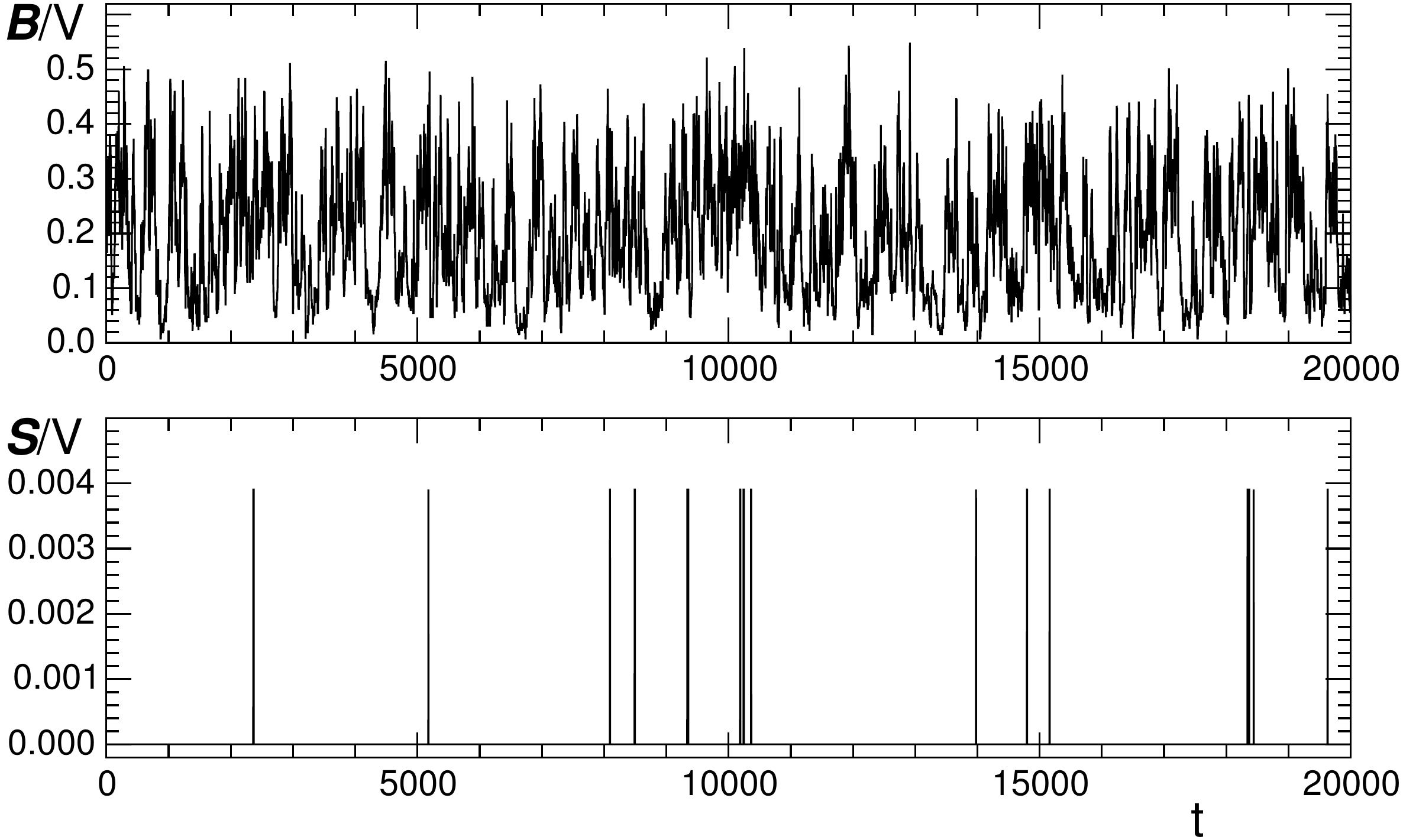}
\vspace*{-2mm}
\caption{Time series for ${\cal B}/V$ and ${\cal S}/V$ at $\kappa = 0.0002$, 
$\mu = 0$ and $\tau = 0.183$ on a $8^3$ lattice, generated with Type-I worms.
The Monte Carlo time on the horizontal
axis is measured in units of $N_{deco} = 20$ worms.} 
\label{small_kappa}
\end{figure}

There are, however, regions in the space of couplings where one finds that the
flux representation and worms are not the optimal approach. In general this is
the case for very small values of  $\kappa$ (typically for $\kappa$ of
$O(10^{-4})$), where the monomer weights are small (in leading order they are
linear in $\kappa$), and thus it is hard to generate monomers. This is
illustrated in Fig.~\ref{small_kappa}, where we show the same time series as in
Fig.~\ref{large_kappa}, but now at $\kappa = 0.0002$, $\mu = 0$ and $\tau =
0.183$, again with $N_{deco} = 20$ worms for decorrelation.  Although the time
series for ${\cal B}/V$ still looks reasonable, the time series for ${\cal S}/V$
shows that monomers are generated only very rarely and very high statistics
would be needed to obtain reliable results.  We stress that this is a property of
the flux representation: As mentioned, the monomer weights are very small at
small $\kappa$, so monomers are indeed expected to be very rare and the flux
representation is not the optimal choice for very small $\kappa$. This
difficulty could be overcome by reweighting the sector of phase space with
small monomer numbers to obtain a higher weight and thus higher frequency in the
Monte Carlo. They could then be weighted back in order to compute expectation values.  In the
present study we did not explore such a reweighting strategy. We also note that
for small $\kappa$ the complex phase problem is very mild and reweighting
techniques were used to successfully simulate the system with the conventional
spin representation \cite{Alford,Kim} in that region of the parameter space also
at small $\mu$.

\subsection{Characteristic quantities}

Let us now come to a quantitative comparison of the Type-I closed worms and
Type-II open worms. For a sensible comparison one has to consider the actual
changes done by the worms: A Type-II open worm inserts two monomers at its
endpoints and connects it with dimers (one segment), while a Type-I closed worm
may consist of several segments of dimer strings connected by large hops between
monomers (it may also consist of only dimer hops without large hops, i.e., only
one closed segment).   We first look at some characteristic quantities in order
to describe the behavior of the worms in different regions of parameter space.
By $n_{os}$ we denote the  average number of  open
segments in Type-I worms, i.e., segments with monomer hops at their endpoints
(if a Type-I worm closes without monomer insertions $n_{os}$ is zero). By $D$ we
denote the average number of dimer steps  in a Type-I or Type-II worm. Another
interesting quantity for the comparison of the two worms is the ratio $r$  of
the number of successful worm starts divided by the number of  all start
attempts. Depending on the parameters, here one expects a drastic difference
between the worms, since the open worms can start only with the insertion of a
monomer. Finally, for the comparison of the computational cost of the two
algorithms we define the cost ratio $c_s$ for the Type-I and Type-II worms as the ratio
of the total number of attempted steps (dimer, monomer and start attempts) to
the number of accepted steps. 

\begin{table}[t!]
\begin{center}
\hspace*{-8mm}
\begin{tabular}{lrllllll}
\hline
Parameters & $V$ & $\;\;\;r$  &  $\;n_{os}/V$   & $\;\;\;D/V$  & $\;\;\;c_s$ \\
\hline
{\bf Set A:}       & $ 6^3$ & 0.166 & 0.122 & 0.383 & 3.81 \\
$\tau = 0.100,$    & $ 8^3$ & 0.166 & 0.122 & 0.380 & 3.80 \\
$\kappa = 0.001,$  & $16^3$ & 0.166 & 0.122 & 0.378 & 3.80 \\
$\mu = 5.9$        & $32^3$ & 0.166 & 0.122 & 0.377 & 3.80 \\
\hline
{\bf Set B:}       & $ 6^3$ & 0.209 & 2.50e-4 & 0.332 & 3.26 \\
$\tau = 0.181,$    & $ 8^3$ & 0.200 & 2.29e-4 & 0.221 & 3.51 \\
$\kappa = 0.001,$  & $16^3$ & 0.190 & 1.82e-4 & 0.066 & 4.81 \\
$\mu = 0.8$        & $32^3$ & 0.188 & 1.44e-4 & 0.031 & 6.32 \\
\hline
{\bf Set C:}       & $ 6^3$ & 0.129 & 0.445 & 0.212 & 3.89 \\
$\tau = 0.025,$    & $ 8^3$ & 0.129 & 0.444 & 0.212 & 3.89 \\
$\kappa = 0.005,$  & $16^3$ & 0.129 & 0.444 & 0.211 & 3.88 \\
$\mu = 5.2$        & $32^3$ & 0.129 & 0.444 & 0.212 & 3.89 \\
\hline
{\bf Set D:}       & $ 6^3$ & 0.173 & 7.59e-4 & 0.126 & 3.99 \\
$\tau = 0.170,$    & $ 8^3$ & 0.170 & 5.66e-4 & 0.059 & 4.64 \\
$\kappa = 0.005,$  & $16^3$ & 0.169 & 4.18e-4 & 0.017 & 7.25 \\
$\mu = 0.2$        & $32^3$ & 0.169 & 4.03e-4 & 0.012 & 8.73 \\
\hline
{\bf Set E:}       & $ 6^3$ & 0.164 & 0.013 & 0.197 & 4.61 \\
$\tau = 0.150,$    & $ 8^3$ & 0.164 & 0.012 & 0.178 & 4.70 \\
$\kappa = 0.010,$  & $16^3$ & 0.164 & 0.012 & 0.166 & 4.77 \\
$\mu = 2.0$        & $32^3$ & 0.164 & 0.012 & 0.164 & 4.78 \\
\hline
\end{tabular}
 
\end{center}
\caption{Characteristic quantities for Type-I closed worms (see the text for their definitions).  
All the simulations use $10^6$ worms for equilibration and $10^7$ measurements
separated by $N_{deco} = 5$ worms for decorrelation. The error is
usually smaller than the last digit we show.}
\label{count_closed}
\end{table}

\begin{table}[t!]
\begin{center}
\begin{tabular}{lrllll}
\hline
Parameters & $V$  & $\;\;\;r$ & $\;\;\;D$  & $\;\;\;c_s$ \\
\hline
{\bf Set A:}       & $ 6^3$ & 0.268 & 2.93 & 3.78 \\ 
$\tau = 0.100,$    & $ 8^3$ & 0.268 & 2.93 & 3.78 \\
$\kappa = 0.001,$  & $16^3$ & 0.268 & 2.93 & 3.78 \\
$\mu = 5.9$        & $32^3$ & 0.268 & 2.93 & 3.78 \\
\hline
{\bf Set B:}       & $ 6^3$ & 0.0014 & 108 & 9.44 \\
$\tau = 0.181,$    & $ 8^3$ & 0.0015 & 163 & 7.17 \\ 
$\kappa = 0.001,$  & $16^3$ & 0.0016 & 195 & 6.42 \\
 $\mu = 0.8$       & $32^3$ & 0.0015 & 179 & 6.77 \\ 
\hline
{\bf Set C:}       & $ 6^3$ & 0.625 & 0.445 & 3.85 \\
$\tau = 0.025,$    & $ 8^3$ & 0.626 & 0.445 & 3.85 \\
$\kappa = 0.005,$  & $16^3$ & 0.625 & 0.445 & 3.85 \\ 
$\mu = 5.2$        & $32^3$ & 0.626 & 0.445 & 3.85 \\ 
\hline
{\bf Set D:}       & $ 6^3$ & 0.0059 & 30.1 & 8.57 \\
$\tau = 0.170,$    & $ 8^3$ & 0.0058 & 29.2 & 8.77 \\
$\kappa = 0.005,$  & $16^3$ & 0.0058 & 28.3 & 8.98 \\
$\mu = 0.2$        & $32^3$ & 0.0058 & 28.3 & 8.98 \\
\hline
{\bf Set E:}       & $ 6^3$ & 0.0482 & 12.5 & 4.70 \\
$\tau = 0.150,$    & $ 8^3$ & 0.0482 & 12.5 & 4.70 \\
$\kappa = 0.010,$  & $16^3$ & 0.0482 & 12.5 & 4.70 \\ 
$\mu = 2.0$        & $32^3$ & 0.0482 & 12.5 & 4.70 \\ 
\hline
\end{tabular}
\end{center}
\caption{Characteristic quantities for Type-II open worms (see the text for their definition).  
All the simulations use $10^6$ worms for equilibration and $10^7$ measurements
separated by $N_{deco} = 500$ worms for decorrelation. The error is
usually smaller than the last digit we show.}
\label{count_open}
\end{table}

In Table \ref{count_closed} we collect the data for Type-I closed worms, and in
Table \ref{count_open} for Type-II worms. The analysis was done for five
different sets of parameters (first column) and for four volumes (second
column). Let us begin with the comparison of the ratio $r$ (number of successful
worm starts  divided by the number of all start attempts). For Type-I closed
worms the starting probability is similar for all parameter sets
(ranging between 0.129 and 0.209). Since here the chance for starting with a
dimer insertion is 6/7, the ratio goes down slightly when dimers are more
costly, i.e., when $\tau$ is small, or when the system is monomer dominated
($\kappa$ and/or $\mu$ large). For the Type-II open worms, $r$ shows a much
larger variation: Here the worm must start with the insertion of a monomer, and
for some parameter sets (Sets B, D and E) these are scarce, such that for the
Type-II worms many start attempts fail. On the other hand, for parameters where
monomers are abundant (Sets A and C), we observe a very high rate of
successful starts. 

For Type-I closed worms the abundance of monomers is also
reflected in the average number $n_{os}$ of open segments: It is small for the
same parameter sets (Sets B, D and E) where the starting ratio $r$ of the
Type-II worms is small, i.e., for regions of parameter space with few monomers.
Concerning the dependence on the volume $V$, we find that $n_{os}$ grows linearly
with the volume, i.e., $n_{os} = c \, V$ perfectly describes the data,
where $c$ is a constant that depends on the parameters $\tau, \kappa$ and $\mu$.

The behavior of the average number $D$ of dimer steps is interesting. For Type-II
open worms we find that $D$ varies only very little with the volume and is correlated with
the abundance of monomers: Since monomers are needed to terminate Type-II
worms, parameter values where monomers are scarce (Sets B, D and E) will lead
to longer worms, i.e., increase $D$. For Type-I closed worms we find that 
$D$ essentially grows linearly with the volume for Sets A, C and E. For the
sets B and D (those with very low density of monomers) we find that $D$
grows slower than linear.
This is correlated with a very small number $n_{os}$ of open segments for these ensembles. 

The cost ratio $c_s$ essentially reflects the behavior of $r$, $n_{os}$ and
$D$:
Whenever monomers are sufficiently abundant, the values of $c_s$ are independent
of the volume and are roughly equal for Type-I and Type-II worms. Discrepancies
are observed only for the sets where monomers are scarce (Sets B and D): In
these cases the closed Type-I worms have the advantage that they can run without
any monomer insertions at all, thus improving their cost ratio. This advantage is,
however, lost when the volume is larger and longer dimer chains emerge. Then
from time to time a monomer insertion takes place and also Type-I worms then
spend quite some time with trying to insert another monomer necessary for
continuation. Thus for larger volumes the cost ratios $c_s$ of Type-I and Type-II worms
roughly agree also for Sets B and D. 
From the assessment of $r$, $n_{os}$, $D$ and $c_s$ we conclude that 
for these characteristic quantities the two
algorithms behave similarly, with the exception that the closed Type-I worms
are more flexible since they work also for vanishing external magnetic fields.

\subsection{Autocorrelation times}
 
The relevant figure of merit for an algorithm is computational effort
for a fixed precision of results.
We therefore analyze the integrated autocorrelation time $\tau_{int}^X$ of  the bulk
observables $X = U, C, P$ and $\chi_p$. 
%
%
Since the sizes of worms vary drastically between the different cases, we
need to normalize the autocorrelation times.
For this purpose we define 
one ``sweep'' as $\tau_0 = 3V/D$ worms, i.e., the average number of worms needed 
to visit every link of the lattice
as the (customary) unit for the integrated autocorrelation times $\tau_{int}^X$.
Obviously one may express $\tau_0$ also in
terms of measurements as $\tau_0 = 3V/(D \, N_{deco})$ measurements. 
In Tables 3 and 4 we give the autocorrelation times in units of $\tau_0$.
In order to obtain a measure of computational effort, the results are multiplied by
the cost ratio $c_s$, in other words we show $\overline{\tau} = 
c_s \, \tau_{meas}/\tau_0$, where $\tau_{meas}$ simply is the unnormalized
autocorrelation time in units of measurements.  
The statistical errors of autocorrelation times were estimated by a jackknife
procedure and are between about 5 and 10 percent 
depending on the parameter values. 
This is sufficient for the subsequent comparison of the two algorithms.

The autocorrelation times are almost independent of volume for most parameter sets,
reflecting the crossover or noncritical nature of correlations.
Even for set B, which is very close to the critical value of $\tau$ at $\kappa=0$, 
autocorrelation times of Type-I are small and increase only very moderately with volume.
Self averaging within a large volume can lead to the decreasing times (at high scale)
for the internal energy $U$ for set B and Type-II.

Comparing the autocorrelation times 
(Tables 3 and 4) shows that
Type-II open worms outperform the Type-I closed worms (i.e., shorter autocorrelation times) 
for those sets where monomers are abundant (Sets A and C), 
while the Type-I worms perform better for Sets B, D, E, where monomers are scarce.
The reason is that Type-II worms always start with the insertion of a monomer,
while Type-I worms insert monomers only with a probability 1/7. Thus Type-II
worms may have a larger portion of monomer insertions compared to dimer steps,
and thus do many monomer changes which is advantageous for Sets A and C. Vice
versa, for the sets with few monomers, the Type-I worms can do many dimers steps
without needing monomer insertions and thus perform better in this situation. 
We remark that some of the disadvantage of the Type-I worms for parameter sets where
monomers are abundant could be ameliated by changing the probability for
offering a monomer change from 1/7 to some larger value.
Remarkably, the relative performance of the two worm types is roughly proportional 
to $r/c_s$, i.e. to the acceptance rates for attempted steps.
%
%

\begin{table}[t!]
\begin{center}
\begin{tabular}{lrrrrrr}
\hline
Parameters & $V$ &  $\overline{\tau}^{U}_{int}$  & $\overline{\tau}^{C}_{int}$  & 
$\overline{\tau}^{P}_{int}$  & $\overline{\tau}^{\chi_P}_{int}$  \\
\hline
{\bf Set A:}      & $ 6^3$ & 31 & 17 & 26 & 15 \\
$\tau = 0.100,$   & $ 8^3$ & 31 & 17 & 27 & 14 \\
$\kappa = 0.001,$ & $16^3$ & 31 & 17 & 26 & 14 \\
$\mu = 5.9$       & $32^3$ & 32 & 18 & 27 & 15 \\
\hline
{\bf Set B:}      & $ 6^3$ & 83  & 30 & 19 & 15 \\
$\tau = 0.181,$   & $ 8^3$ & 99  & 35 & 23 & 18 \\
$\kappa = 0.001,$ & $16^3$ & 151 & 72 & 78 & 21 \\
$\mu = 0.8$       & $32^3$ & 156 & 73 & 89 & 32 \\
\hline
{\bf Set C:}      & $ 6^3$ & 7.8 & 5.0 & 6.7 & 4.9 \\
$\tau = 0.025,$   & $ 8^3$ & 7.9 & 5.1 & 7.5 & 4.8 \\
$\kappa = 0.005,$ & $16^3$ & 7.7 & 5.2 & 8.0 & 5.0 \\
$\mu = 5.2$       & $32^3$ & 7.7 & 5.2 & 7.9 & 5.1 \\
\hline
{\bf Set D:}      & $ 6^3$ & 27 & 16 & 13 & 6.9 \\
$\tau = 0.170,$   & $ 8^3$ & 25 & 13 & 13 & 6.1 \\
$\kappa = 0.005,$ & $16^3$ & 32 & 14 & 11 & 5.9 \\
$\mu = 0.2$       & $32^3$ & 54 & 28 & 11 & 6.2 \\
\hline 
{\bf Set E:}      & $ 6^3$ & 48 & 18 & 51 & 20 \\
$\tau = 0.150,$   & $ 8^3$ & 82 & 38 & 86 & 42 \\
$\kappa = 0.010,$ & $16^3$ & 79 & 42 & 82 & 43 \\
$\mu = 2.0$       & $32^3$ & 77 & 39 & 80 & 43 \\
\hline
\end{tabular}
\end{center}
\caption{Type-I closed worms: Autocorrelation times of the bulk observables at
different parameters.  We show the autocorrelation times $\overline{\tau}$ 
in units of $\tau_0$ and multiplied by the cost ratio $c_s$ as discussed in the text.}
\label{auto-cw}
\end{table}

\begin{table}[t!]
\begin{center}
\begin{tabular}{lrrrrrr}
\hline
Parameters & $\;\;\; V$  & $\overline{\tau}^{U}_{int}$  & $\overline{\tau}^{C}_{int}$  & 
$\overline{\tau}^{P}_{int}$  & $\overline{\tau}^{\chi_P}_{int}$  \\
\hline
{\bf Set A:}      & $6^3$  & 14 & 6.8 & 12 & 5.1 \\
$\tau = 0.100,$   & $8^3$  & 15 & 7.2 & 13 & 5.0 \\
$\kappa = 0.001,$ & $16^3$ & 15 & 6.9 & 13 & 5.1 \\
$\mu = 5.9$       & $32^3$ & 14 & 7.0 & 11 & 4.9  \\
\hline
{\bf Set B:}      & $6^3$  & 35000 & 13000 & 1300  & 1200 \\
$\tau = 0.181,$   & $8^3$  & 31000 & 13000 & 7800  & 2700 \\
$\kappa = 0.001,$ & $16^3$ & $>$25000 & 15000 & 12000 & 3900\\
$\mu = 0.8$       & $32^3$ & $>$19000 & $>$11000 & $>$11000 & $>$3100\\
\hline
{\bf Set C:}      & $6^3$  & 1.2 & 0.53 & 1.1 & 0.47 \\
$\tau = 0.025,$   & $8^3$  & 1.1 & 0.50 & 1.0 & 0.46 \\
$\kappa = 0.005,$ & $16^3$ & 1.2 & 0.53 & 1.1 & 0.47 \\
$\mu = 5.2$       & $32^3$ & 1.2 & 0.53 & 1.4 & 0.49 \\
\hline
{\bf Set D:}      & $6^3$  & 2000 & 990 & 180 & 80 \\
$\tau = 0.170,$   & $8^3$  & 1400 & 740 & 210 & 100 \\
$\kappa = 0.005,$ & $16^3$ & 1200 & 580 & 230 & 100\\
$\mu = 0.2$       & $32^3$ & 940 & 540 & 220 & 85 \\
\hline
{\bf Set E:}      & $6^3$  & 240 & 120 & 240 & 120\\
$\tau = 0.150,$   & $8^3$  & 230 & 110 & 240 & 120\\
$\kappa = 0.010,$ & $16^3$ & 230 & 120 & 240 & 110\\
$\mu = 2.0$       & $32^3$ & 230 & 110 & 240 & 110\\
\hline
\end{tabular}
\end{center}
\caption{Type-II open worms: Autocorrelation times of the bulk observables at
different parameters.  We show the autocorrelation times $\overline{\tau}$ 
in units of $\tau_0$ and multiplied by the cost ratio $c_s$ 
as discussed in the text.  For cases marked with ``$>$'', 
only a lower bound could be determined.}
\label{auto-ow}
\end{table}

\section{Summary and discussion}
In this article we presented in detail two worm algorithms that were used in a
recent  QCD-inspired study of the 3-state Potts model with external field and
chemical potential  in three dimensions \cite{effcenter} and analyzed their
performance. The algorithms are based on a flux representation of the model
which can be obtained using high temperature expansion techniques. The nearest
neighbor term leads to the conventional dimer-based closed contours, while the
magnetic terms allow violation of the constraint through the insertion of
monomers (similar structures will appear whenever one considers spin systems 
that are coupled to an external field). Furthermore, the chemical potential
gives a different weight to monomers and antimonomers. Thus the model provides a
rather general testbed for various algorithmic ideas based on the worm concept
\cite{worm}. 

We explored two types of generalized worm algorithms that differ in their
treatment of monomers. The Type-I closed worms start with dimers or monomers and
allow for dimer insertions or the insertion of monomers which are then followed
by a random hop to another position where a second monomer is inserted. These
closed worms finish when the starting position is reached again.
On the other hand the Type-II open worms create chains of dimers that start and end with the 
insertion of monomers. It is important to note that the open worms are
restricted to simulations with non-vanishing external magnetic field. 
 
It was carefully checked that the two worm algorithms produce correct results,
by comparison with an exact evaluation on small volumes, a cross-check with a
standard Metropolis simulation (at vanishing chemical potential), and a
systematic comparison of the results from the two algorithms at various
parameter values.  

In order to evaluate the performance of the two algorithms we first compared various
characteristic quantities, namely the starting probability $r$, the number of
open segments $n_{os}$ in Type-I worms, the average number of dimer steps
$D$ of the worms, and their cost  ratio $c_s$. 
We then presented the results of an analysis of the autocorrelation times, 
suitably scaled to reflect computational effort.
The overall assessment shows that the
Type-I closed worms are more flexible: They can be used also at vanishing
external field and in general were found to perform better for parameter sets
where monomers are scarce. Type-II open worms perform considerably better in monomer dominated 
regions of the parameter space. 
However, we expect that changing the ratio of 
monomer to dimer steps in Type-I closed worms would improve the latter 
for situations where monomers are abundant. 

The authors expect that generalizations of the worm concept will see a lot of
attention in the future: Matter fields (bosonic and fermionic) give rise to
closed loops of flux on a lattice, and worm algorithms are a natural approach
to update such a system. Of course in more realistic models the structure of the
fluxes is more involved (see, e.g., the generalized effective Polyakov loop
model \cite{cg}) and new strategies need to be found. We expect that the current
paper is useful for such further developments.  

\newpage
\section*{Acknowledgments}
We thank Philippe De Forcrand, Daniel G\"oschl, Christian Lang, Gundolf Haase and Manfred Liebmann
for fruitful discussions at various stages of this work. Y.~Delgado
is supported by the FWF Doktoratskolleg {\sl Hadrons in Vacuum, Nuclei and
Stars} (DK W1203-N08) and by the Research Executive Agency (REA) of the European Union 
under Grant Agreement number PITN-GA-2009-238353 (ITN STRONGnet).

\end{document}